\documentclass[conference]{IEEEtran}
\IEEEoverridecommandlockouts
\usepackage{cite}
\usepackage{amsmath,amssymb,amsfonts}
\usepackage{algorithmic}
\usepackage{graphicx}
\usepackage{textcomp}
\usepackage{xcolor}
\def\BibTeX{{\rm B\kern-.05em{\sc i\kern-.025em b}\kern-.08em
    T\kern-.1667em\lower.7ex\hbox{E}\kern-.125emX}}
\begin{document}

\title{A New Communication Protocol with Self Error Correction}

\author{\IEEEauthorblockN{Ye Tianyi}
\IEEEauthorblockA{\textit{School of Mechanical, Electrical and Information Engineering} \\
\textit{Shandong University}\\
admin@nulladev.cn}
}

\maketitle

\begin{abstract}
Communication in poor network environment is always a difficult problem, since troubles such as bit errors and packet loss may often occur. It is generally believed that it is impossible to transmit data both accurately and efficiently in this case. However, this paper provides a method to transmit data efficiently on the line where bit error may occur by utilizing Hamming code principle. If the sender adds a small amount of redundant data to the data to be sent, the receiver can self-correct them when an error is detected. This approach takes advantage of the value of packets with errors, which should have been discarded, reduce the number of re-transmissions and improve transmission efficiency. Based on this method, this paper designs a custom protocol which works in the data link layer and network layer. Finally, this paper verifies the protocol through mathematical simulation.
\end{abstract}


\section{Introduction}
Imagine a scene: on a stormy night, you are anxiously texting your family to pick you up. However, due to the interference of lightning, the mobile phone signal becomes very poor and finally your family only received some “bad” packets that do not pass the verification. Is that reasonable? If the network is very poor, is there any other way besides requiring the sender to re-transmit? Can we get some information from the code with errors? These series of questions lead us to Hamming code. 

Hamming code is a series of linear error correction codes invented by R. W. Hamming in 1950, which can detect, locate and correct errors with a relatively small amount of redundancy when the error rate is low. Because this condition is met, this algorithm is widely used in RAM, called ECC memory \cite{b1}.

According to the principle of Hamming code, this paper proposes a new frame structure in data link layer (yes, we shouldn't call it Ethernet Frame anymore!), and also a communication protocol using this structure in network layer. The biggest difference between the new structure and Ethernet frames specified in IEEE 802.3 standard is that the frame check sequence (FCS, 32-bit CRC) at the end of each frame is replaced by Hamming code with similar functions. Meanwhile in the transmission protocol, special transmission logics are also designed in order to match a hypothetical poor network conditions.

According to our assumption, the protocol is applicable to the low communication speed but error prone situations, such as satellite communication. In actual satellite communication, the speed varies greatly depending on the communication distance. According to the data released by NASA, the communication speed between the Curiosity and its artificial satellite can reach 128,000 or 256,000 bits per second, while the transmission speed from the artificial satellite to ground is only 500 bits per second to 32,000 bits per second \cite{b2}. Although the project does not give the exact bit error rate, in other Earth-Mars communication links, the bit error rate usually only needs to reach 10\textsuperscript{-5} \cite{b3}. If the protocol can be used in this situation, it can reduce the number of re-transmissions of error packets and benefit from expensive communication costs. 

\section{Related Works}

Although has been widely used in various hardware such as RAMs, Hamming code is still rarely used in network communication up to now. The reason, we think, is that the old Internet communication standards limit the change of protocols. In fact, the structure of Ethernet frames has not been modified greatly since the IEEE 802.3 standard was proposed in 1983. As more and more people use this standard, it also becomes more and more difficult to modify it.

Through reviewing the historical papers, we found that some published articles have mentioned the detection and correction of communication errors through Hamming code, such as \cite{b4}, \cite{b5}, \cite{b6}. However, these articles used almost all the space to introduce the principle of Hamming code, and did not design a reasonable protocol to use Hamming code. In addition, in the existing network architecture, packets with bit errors will be directly discarded because they cannot pass the FCS check. Therefore, the effect of Hamming code is difficult to be tested in the real world unless special hardware is first manufactured. Therefore, we have observed that Hamming code has been used in smart grid \cite{b7}, CAN \cite{b8} and tele robotic system \cite{b9}, all of which belong to a small private system rather than a wider protocol.

At the same time, some characteristics of Hamming code also limit the use of the algorithm. For example, the additional calculation cost for sending and receiving data packets, the completely immutable packet size and the number of checksum, etc.

\section{Methodology}

\subsection{Methodology of Hamming code}

The basic ideas of Hamming code on error detection and correction are as follows:
\begin{itemize}
\item The effective information is divided into several groups according to a certain rule, each group is arranged with a check bit, which can obtain the specific check code by XOR operation.
\item At the receiving end, the XOR operation is also used to determine the correction of the verification results of each group and to obtain the specific error bit by observing the error correction group or the joint check bit of multiple error check groups.
\item Correct the error bit by inverting it.
\end{itemize}

For example, we have a block of 16 bits and number these bits from 0 up to 15. Only 11 of these bits will be used to store data, while 5 of the other positions will be reserved as a kind of redundancy. 4 of these redundancies will sit in positions which are powers of 2(Show in the Fig. 1 and will be explained later) and will be used for Parity Checks. The 0\textsuperscript{th} bit is also a redundancy and will be used to detect whether there are two or more errors in the block (This part will be explained in detail later).

\begin{figure}[htbp]
\centerline{\includegraphics[scale=0.375]{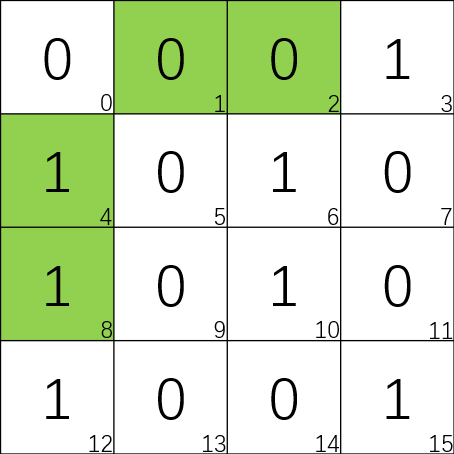}}
\caption{Four redundancies used for Parity Checks.}
\label{fig1}
\end{figure}

In short, the Parity Check means that we calculate the total numbers of “1” in a given sets. And this is very meaningful to help us find error in the 16 bits (or bigger) block.

Firstly, we consider the situation that there is only one error in the block. If we apply some parity checks to certain selected subsets, it will help us pin down the location of any single-big error.

\begin{figure}[htbp]
\centerline{\includegraphics[scale=0.25]{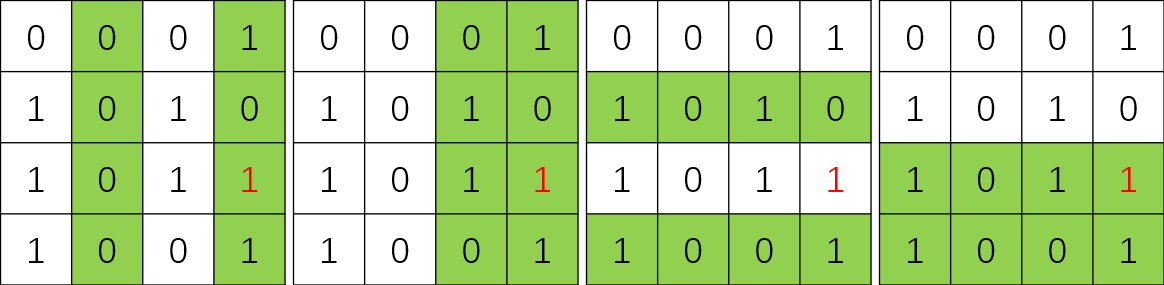}}
\caption{Four certain selected subsets for Parity Checks (Q1-Q4).}
\label{fig2}
\end{figure}

For example, let’s do a parity check, such as subset Q1 displayed in Fig. 2. The result of the total count of “1” in this subset is odd, which means that there must be one error in these eight bits. And then we continue doing the parity checks on each subset, such as subsets Q2, Q3 and Q4 displayed in Fig. 2. Each time we finish a parity check, we will narrow the possible space for error. Finally, we will find the error located in the 11th bit, which means that we can not only realize there exists error in the block but also locate the error and correct it. And the first bit in each of the four subsets will be used as a parity bit respectively. At the same time, we find the position of these bits are 1\textsuperscript{st}, 2\textsuperscript{nd}, 4\textsuperscript{th} and 8\textsuperscript{th} bit, which are powers of 2. In fact, our goal is to protect the message bits, while the parity bits are also protected by the way. This method can be promoted to a bigger block if we know the sum of bits is less than or equal to the nearest n\textsuperscript{th} power of 2. And anyone can do n times parity check, such as we show in Fig. 2 to locate the position of error.

If there are more than one error exist in the block, we can also detect it with the help of 0th bit. Firstly, we should do a parity check for the whole block, if the result is odd, that means there are odd number of errors in the block. Otherwise, there may be no error or even number of errors in the block. Then we do parity checks like Fig. 2 again, and if any result of these parity check is odd, that means there must be even number of errors in the whole block. Only when all results of parity checks are even, we can judge that there is no error in the block.

When we apply Hamming codes in transmission like the 16 bits block, the sender is responsible for setting these special parity bits. And the receiver is responsible for performing some checks on the block and then they can correct error or at least they can tell there exist error in the block.

\subsection{Programming implementation of Hamming code}

After we have introduced the methodology of Hamming code, one may believe that we need to write an algorithm to keep track of all possible error locations and cut that group in half with each check. However, the implementation is far simpler than one can imagine.

\begin{figure}[htbp]
\centerline{\includegraphics[scale=0.25]{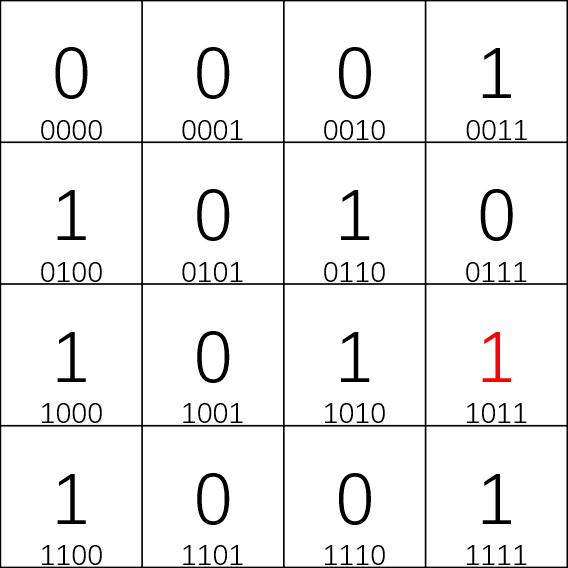}}
\caption{Use binary number to label the 16 bits block.}
\label{fig3}
\end{figure}

Again, we utilize the 16 bits block to present the explanation. At the beginning, we should label these 16 positions in binary from 0000 up to 1111(shown in the Fig. 3). Then we only pay attention to the position, where that final bit is 1 and we can get the first of four parity groups (shown in the Fig. 2), which means that you can interpret our previous check as asking: “If there is an error, is the final bit in the position of that error is ‘1’?” Similarly, if focusing on the third bit, the second bit and the first bit, one will find the similar conclusion, which means that the location of the error has been directly told to us in binary and this conclusion can be promoted to be a bigger block. If it takes more bits to label each position, such as 64 bits block need 6 bits to label, then each of these bits represents one of the parity groups that we need to check. Parity bits are sitting in positions that are n powers of two, because these positions’ binary representation possesses just one bit is “1” and the other bits are all “0”. Furthermore, these parity bits sit in only one of all parity groups. This also appears in larger examples, no matter how big it is, each parity bit conveniently falls on the only one of the parity groups.

After we realize that the parity checks are just a clever method to discover the position of an error in binary, we can understand the Hamming codes in a different way, which is far simpler and elegant and can be implemented with a single line of code. The method is based on the XOR function and XOR stands for “exclusive or”. When one takes the XOR of two bits, it will return 0 only when both the two bits are either “0” or “1”. Otherwise, it would return 1. In other words, XOR can represent the parity of these two bits. Moreover, there is an addition of these two numbers and then mod 2. This provides us with a beautiful method to think about the multiple parity checks from Hamming code algorithm, since all calculations are simplified to one single operation - XOR. Specifically, write down the 16 positions in binary and highlight the positions that are 1(show in Fig. 4). And then collect these positions’ binary label together into one column and take the XOR (show in the Fig. 5). You will coincidentally find the result is the label of the position where the error sit. The last column, for example, is counting all of the positions whose last bit is 1. But we have already limited these positions to the highlighted ones, so we count how many highlighted positions are from the first parity group. Similarly, we count the total number of “1” in 1\textsuperscript{st}, 2\textsuperscript{nd} and 3\textsuperscript{rd} bit respectively and discover how many highlighted positions come from the corresponding parity groups. In fact, the sender is responsible for toggling some of the special parity bits to make sure the XOR sum of these highlighted positions is 0000. Obviously, if one error appears in the 16 bits block, the binary label of the error will be either added (“0”→“1”) or removed (“1”→“0”) from the XOR calculation of positions that are 1. Whether the error is “0”→“1” or “1”→“0”, the final XOR result will always be the binary label of the position where the error appears.

\begin{figure}[htbp]
\centerline{\includegraphics[scale=0.25]{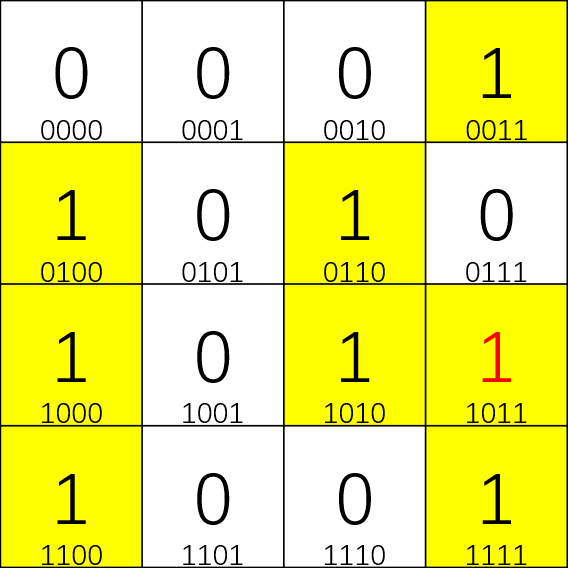}}
\caption{Highlight the positions that are 1 in the 16 bits block.}
\label{fig4}
\end{figure}

\begin{figure}[htbp]
\centerline{\includegraphics[scale=0.5]{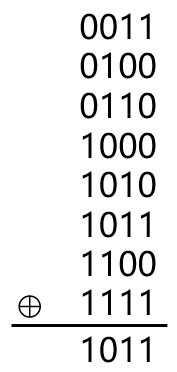}}
\caption{The result of XOR operation.}
\label{fig5}
\end{figure}

\section{Design}

\subsection{Packet design}

\textbf{Packet Size.} As we all know, data is usually stored and transmitted in bytes rather than bits in electronic devices. To avoid the situation that frames containing a payload of non-integer bytes, the number of redundant bits should be a multiple of 8. Therefore, the total size of each part of the packet is fixed, which can only be 2\textsuperscript{7}, 2\textsuperscript{15}, or 2\textsuperscript{23} bits. For convenience, we call it small components, medium components and large components below.

For a small component, it has a total size of 2\textsuperscript{7} = 128bits = 16bytes. This is too small to be used for any data transfer. Therefore, we think it is appropriate to carry control information, such as a protocol header. 

For a medium component, it has a total size of 2\textsuperscript{15} = 32768bits = 4096bytes = 4Kb. This size is similar to the size of a typical IP packet shard (1500 bytes), so it can also be used similarly. Therefore, we decided to use this kind of components as payloads.

For a large component, it has a total size of 2\textsuperscript{23} = 8388608bits = 1048576bytes = 1Mb. Since we assume that the communication environment is extremely poor, it may be difficult to recover data with Hamming code for this size. Although thinking this kind of component was useless, we finally designed it anyway.

\textbf{Packet Structure.} According to above, we have already obtained the basic structure above layer 2 of a packet. In order to be compatible with the existing physical layer protocol, we decided not to change the structure of layer 1 in the protocol. This means that we retain the seven-bytes preamble (0x55 0x55 0x55 0x55 0x55 0x55 0x55) and one-byte start frame delimiter (SFD, 0xD5) added by the physical layer. Thus, the total structure of a packet is shown in Table 1.

\begin{table}[htbp]
\caption{Total structure of a packet}
\begin{center}
\begin{tabular}{|c|c|c|c|c|c|}
\hline
Layer & Preamble & SFD & Layer 2 Header & Payload & FCS \\
\hline
 & 7 Bytes & 1 Byte & 16 Bytes & Variable & No \\
\hline
1 & \multicolumn{4}{c|}{yes} & \\
\hline
2 & & & \multicolumn{2}{c|}{yes} & \\
\hline
\end{tabular}
\label{tab1}
\end{center}
\end{table}

\textbf{Layer 2 Header.} A regular Ethernet frame header features destination and source MAC addresses (6 bytes each), EtherType fields (2 bytes), and optional IEEE 802.1Q tags or IEEE 802.11d tags (4 bytes). Among them, the EtherType field can be used for two different purposes. When it is 1500 or less, it indicates the number of bytes of the payload. When it is 1536 and above, it represents the protocol number encapsulated in the frame payload.

In order to ensure a certain degree of similarity with the existing network, we decide to retain the mac address system in our protocol. However, since the number of bytes of the packet is fixed in our protocol, we only need one byte to represent the packet type. In other words, the protocol number determines the length of the payload. For example, maybe we can define the protocol number of an ordinary communication packet (16 + 4096 bytes, see below) as 0x01, and the protocol number of an ACK (no payload) as 0x02, etc.

In addition to the above, we also need a total of 1 byte (intermittent) to place Hamming code information. Since the total size must be 16 bytes, the last 2 bytes are temporarily useless. We set them at 0 for future use. Thus, the structure of the header is shown in Table 2.

\begin{table}[htbp]
\caption{Structure of the header}
\begin{center}
\begin{tabular}{|c|c|c|c|c|}
\hline
dst MAC & src MAC & EtherType & Hamming code & Preserved \\
\hline
6 Bytes & 6 Bytes & 1 Byte & 1 Byte$^{\mathrm{a}}$ & 2 Bytes \\
\hline
\end{tabular}
\end{center}
$^{\mathrm{a}}$intermittent
\label{tab2}
\end{table}

\textbf{Payload.} The structure of payload in each protocol is different. This paper only designs the structure of common communication packets, and other types need to be improved.

In general, this is a structure similar to IP data gram, which also contains header and payload. In order to ensure a certain degree of similarity with the existing network, we decided to retain the IP address system in the protocol. However, since our header is fixed to be only 16 bytes long (including 1 byte Hamming code), the length of ipv6 is obviously too long for this protocol. Therefore, this protocol only supports ipv4. In the future, if the protocol is extended to support ipv6, the header and payload should be placed together in a 4096-byte medium component.

According to the experience of predecessors, this protocol retains some formats in the IP datagram, including ID, frag flag, offset, TTL, etc., and deletes fields that are useless for this protocol, such as the length and the checksum. After all the necessary fields are added, there is a space of 2 bytes left. We set its initial value to 0. If the Hamming code is used for self-correction during transmission, the error location will be recorded here. The final structure of this protocol is shown in Table 3 and 4.

\begin{table}[htbp]
\caption{Structure of the payload header}
\begin{center}
\begin{tabular}{|c|c|c|c|c|}
\hline
src IP & dst IP & ID & flag & offset \\
\hline
4 Bytes & 4 Bytes & 2 Bytes & 3 bits & 13 bits \\
\hline
TTL & \multicolumn{2}{c|}{Hamming code} & Err pos & Total \\
\hline
1 Byte & \multicolumn{2}{c|}{1 Byte$^{\mathrm{a}}$} & 2 Bytes & 16 Bytes \\
\hline
\end{tabular}
\end{center}
$^{\mathrm{a}}$intermittent
\label{tab3}
\end{table}

\begin{table}[htbp]
\caption{Structure of the payload}
\begin{center}
\begin{tabular}{|c|c|c|}
\hline
content & Hamming code & Total \\
\hline
4094 Bytes & 2 Bytes$^{\mathrm{a}}$ & 4096 Bytes \\
\hline
\end{tabular}
\end{center}
$^{\mathrm{a}}$intermittent
\label{tab4}
\end{table}

\subsection{Strategy of self-correction}

For each node on the transmission link, when it receives a packet, it will verify the Hamming code of the packet, just as they usually check the FCS of ordinary IP packets. Then, if no error is found, the node will change the values like TTL, repackage and forward as it does for ordinary IP packets. If only one error is detected, the node will firstly fix it, then modify the value of error position in header, and finally forward it as usual. If more than two errors are detected, the packet is considered irreparable. The node will discard the packet and reply a NAK to the sender.

For the receiver of transmission, if it receives a packet with no Hamming code detection error and the value of error position in the header is 0, it will reply an ACK to the sender. If it detects only one error, or the value of error position in the header is not 0, It will calculate the hash value of the payload and check with the sender through a packet with the receiving packet ID. If more than two errors are detected, the packet is considered irreparable. The receiver will discard the packet and reply a NAK to the sender.

\subsection{Strategy of packet loss}

In addition to generate bit errors, a poor transmission environment also loses packets frequently. In order to deal with this problem, we still need to decide the strategy of packet loss. 

Since the applicable environment of this protocol is in a very poor network, we decide not to force the protocol to transmit multiple packets at once like TCP and just transmit one by one. In other words, we send a packet repeatedly until the receiver replies with an ACK or after a long timeout.

\section{Evaluation}

Since the protocol conflicts with the firmware of the existing Internet infrastructure, and we cannot simulate a real scenario where bit error often occurs, tests on actual equipment is abandoned. Instead, this paper use calculation on mathematical models to simulate this process.

\subsection{Proportion of payload}

In our protocol, a typical data transmission packet contains an 8-byte Layer 1 protocol header, a 16-byte Layer 2 protocol header, a 16-byte Layer 3 protocol header, a 4094-byte payload, and a 2-byte payload checksum. The ratio of payload to total message length is 4094/4136 = 99.0\%.
For comparison, a typical IP fragment is 1500 bytes long, of which only 1480 bytes are payload. In addition, it also contains an 8-byte Layer 1 protocol header, a 14-byte Layer 2 protocol header, and a 4-byte FCS checksum. The ratio of payload to total message length is 1480/1526 = 97.0\%. Although the gap is small, this protocol still has a little advantage in Proportion of payload.

\subsection{Waste caused by discarded packets}

We assume that packets have an average size of L, and the error probability of each bit in the packet is a fixed value p, then the total number of errors in a packet conforms to a binomial distribution B (L, p). Since L is usually very large while p very small, the scenario can be further simplified as a Poisson distribution P (pL).

Assume that the protocol works in the Earth-Mars communication link mentioned above, the value of p is 10\textsuperscript{-5}. In our protocol, the value of L, or the length of the payload is fixed to 4136 Bytes = 33088 bits. Since there are three independent self-correcting components, a packet can correct up to three bit errors. However, since the length of the payload is greater than the length of the protocol header, for convenience of calculation, we assume that all errors occur in the payload, which means at most one error can be corrected. Under such conditions, we have:

P (error = 0) = 0.718 (no error, cost = 0)

P (error = 1) = 0.238 (self-correct, but need to check the hash with the sender. A packet with 16 Bytes payload will be used, cost = 8 + 16 + 16 = 40 Bytes)

P (error $\geq$ 2) = 0.044 (need to re-transmit, but the cost of a second error in the re-transmission process can be ignored because of the order of magnitude, so cost = 4136 Bytes)

Therefore, the mathematical expectation of cost is 191.5 Bytes / packet. Since each packet contains 4094 bytes of payload, the total cost required to transmit 1Mb (1048576 Bytes) is 49049 Bytes.

In contrast, a typical IP fragment is 1526 Bytes = 12208 bits long. Under such conditions, we have:

P (error = 0) = 0.885 (no error, cost = 0)

P (error $\geq$ 1) = 0.115 (need to re-transmit, considering that a second error occurs again during re-transmission, cost = 1526 / (1 – 0.115) = 1724 Bytes)

Therefore, the mathematical expectation of cost is 198.3 Bytes / packet. Since each packet contains 1480 bytes of payload, the total cost required to transmit 1Mb (1048576 Bytes) is 140491 Bytes.

From the above Earth-Mars simulation communication scenario, we can see that the cost of this protocol is only about 35\% of that of traditional protocols.

\subsection{Disadvantages}

\textbf{Computing burden of network equipment.} Since Hamming code is used everywhere, equipment using this protocol must perform a lot of XOR operations when receiving and sending packets, which can reach 15 * 32768 = 491520 times when receiving or sending a packet. If the transmission speed reaches 1Mb/s, about 256 packets will be processed per second, which will burden each device on the link with 491520 * 256 * 2 = 2.5*10\textsuperscript{8} operations per second. This demand of computing speed is of course nothing for computers, but it is a disaster for network devices with small memory. Therefore, in the case of high communication speed, the protocol brings high hardware requirements and high power consumption to communication devices.

\textbf{Waste of remaining size.} Since the packet size is fixed, the last packet in a data transmission process is usually not full. After receiving the EOF of this transmission, 4094/2 bytes of space are wasted on average. Optional solutions include: using a mutation protocol with Hamming code of non-integer bytes to reduce waste (such as using payloads sized 1024 or 2048 bytes).

\end{document}